# Terahertz-Magnetic-Field Induced Ultrafast Faraday Rotation of Molecular Liquids

Vasileios Balos, Genaro Bierhance, Martin Wolf and Mohsen Sajadi[*]

*Fritz Haber Institute of the Max Planck Society, Berlin, Germany*

*sajadi@fhi-berlin.mpg.de

**ABSTRACT** Rotation of the plane of the polarization of light in the presence of a magnetic-field, known as the Faraday rotation, is a consequence of the electromagnetic nature of light and has been utilized in many optical devices. Current efforts aim to realize the ultrafast Faraday rotation on a sub-picosecond time scale. Thereby, the Faraday medium should allow an ultrafast process by which in the presence of an ultrashort intense magnetic-field, the light polarization rotates. We meet these criteria by applying an intense single cycle THz magnetic-field to simple molecular liquids and demonstrate the rotation of the plane of polarization of an optical pulse traversing the liquids on a sub-picosecond time scale. The effect is attributed to the deflection of an optically induced instantaneous electric polarization under the influence the THz magnetic-field. The resolved Faraday rotation scales linearly with the THz magnetic-field and quadratically with the molecular polarizability.

When a linearly polarized electromagnetic (EM) wave, along with a magnetic-field, propagates through an optically transparent medium its plane of polarization rotates. This effect is termed Faraday rotation and has played an important role in elucidating the EM nature of light [1]. It has also been extensively used in many devices including optical switches [2], optical communication systems [3], quantum memories [4], nuclear magnetic resonance spectrometers [5] and light modulators [6]. Further applications of Faraday rotators rely on the development in two main areas, introducing new Faraday media or exploring new features for existing materials and increasing the modulation speed of Faraday rotators.

So far there has been great progress in finding new materials with large Faraday rotation, such as nitride semiconductor alloys [7,8], two-dimensional electron gases [9], graphene [10] and organic molecules [11]. However, a sub-picosecond light modulation requires an inherently ultrafast microscopic process whereupon the light polarization rotates [12]. For example, in semiconductors the modulation rate is limited by the mobility of the charge carriers and the relaxation processes in the bulk material [13]. Moreover, ultrafast magnetic switching with terahertz (THz) clock rate is required, a goal that is far to reach with current magnetic pulses with temporal duration of about hundreds of picoseconds [14,15].

To increase the modulation speed of the Faraday rotators, we take advantage of the magnetic-field of single cycle THz EM pulses. The potential of THz pulses has so far been explored to observe [16] and in some cases control the low-energy elementary excitations in liquids [17,18], solids [19], and gases [20] on ultrafast timescale. In these examples, the THz *electric-field* interacts with the IR-active dipolar excitations of the samples, while the THz *magnetic-field* interaction is neglected because of its much weaker interaction energy. However, recent studies have shown the great potential of the THz magnetic fields for ultrafast Faraday rotation via their direct coupling to the magnons in antiferromagnetic and paramagnetic materials [24,25]. Here, we aim to diversify the realm of ultrafast Faraday rotators beyond the spin interactions and demonstrate the THz magnetic-field-induced ultrafast Faraday rotation in molecular liquids at ambient conditions. We show that in liquids, an instantaneous polarization induced by a femtosecond optical pulse acts as an ultrashort current burst, which in the presence of a THz magnetic-field, rotates the polarization of the optical pulse. This process in highly polarizable liquids, such as long chain alcohols and alkanes, is more efficient than that of the standard magneto-optic media with giant Verdet constant such as terbium gallium garnet (TGG) [26].

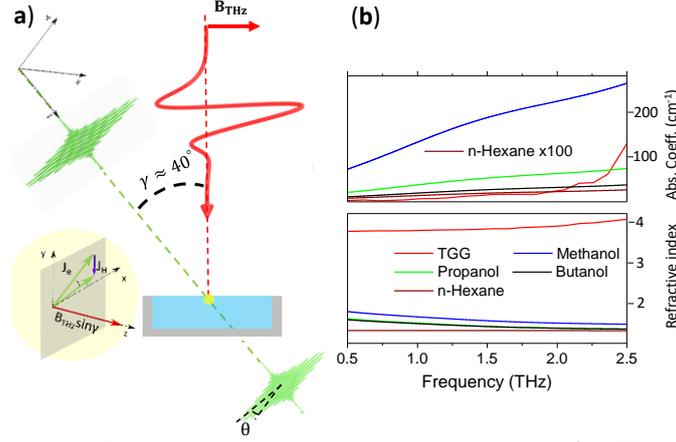

**Figure 1**. a, A femtosecond optical pulse is mixed under a wide angle of ~40° with a single cycle linearly polarized THz pulse on liquids. The optically induced instantaneous polarization, acts as a current burst, $J_e(t)$. This current, is bent by the THz magnetic-field in the plane perpendicular to optical pulse propagation direction. The deflected component of the current, i.e. $J_H(t)$ causes the rotation of the optical pulse polarization. The liquids are held in an open-top bucket. In this configuration, the THz pulse impinges the sample vertically. b, THz absorption and refractive index of some of the samples used in this study. For TGG crystal, data is obtained from our THz time-domain spectrometer. The refractive index and absorption coefficient of methanol [21], butanol, propanol [22] and n-hexane [23] are calculated from their dielectric permittivity and loss spectra of the cited references.

Our experimental setup follows the geometry, introduced by Hoffmann et al. in their pioneering THz Kerr effect work [17]. As shown in Fig. 1a, two EM waves with central frequencies ~1THz (300 **μm**) and ~374THz (0.8 **μm**) and variable time-delay are incident on the samples with a wide angle of $\gamma \approx$ 40°. To avoid contamination of the measured signals by a window response of a cuvette, we hold the liquids in an open-top bucket whose bottom side is transparent at 800 nm and the THz pulse propagates vertically towards the samples.

The optical probe pulse has ~2 nJ energy and ~10fs temporal duration. The THz pulse is generated by the rectification of a 800 nm optical pulse (60 fs, 5 **mJ**, 1 kHz repetition rate) in a LiNbO$_3$ crystal using the tilted-pulse-front technique [27,28]. The single-cycle THz pulse is phase-locked, linearly polarized and when focused to a diameter of ~0.6 mm produces an electric-field $E(t) \approx 2$ MV/cm corresponding to a magnetic-field of $B(t) = E(t)/c \approx 0.7$ T, where $c$ is the speed of light.

The resulting transient effect is obtained by measuring the polarization change of the optical pulse using a set of balanced photodiodes. Due to the THz pump-induced birefringence, the linear polarization of the traversing optical probe pulse with wavelength $\lambda$, through a sample with thickness **L** acquires rotation and ellipticity. The rotation (ellipticity) of polarization $\Delta\theta$ ($\Delta\phi$) is proportional to the phase difference of the left vs right handed circularly (parallel vs perpendicular linearly) polarized components of the probe pulse [29,30]

$$\Delta\theta = \frac{\lambda}{2\pi L}(n_L - n_R) \tag{1a}$$

$$\Delta\varphi = \frac{\lambda}{2\pi L}(n_\parallel - n_\perp) \tag{1b}$$

Where $n_L$ and $n_R$ ($n_\parallel$ and $n_\perp$) are respectively the refractive indices of the material for the circularly polarized left and right handed (linearly polarized parallel and perpendicular) components of the optical probe pulse. The $\Delta\theta$ ($\Delta\phi$) is detected with a combination of a half-wave plate (quarter-wave plate) and a Wollaston prism which splits the incoming beam into two perpendicularly polarized beams with power $P_\parallel$ and $P_\perp$. In the limit $|\Delta\theta|$ and $|\Delta\phi| \ll 1$, the normalized difference $P_\parallel - P_\perp \propto \Delta\theta(\Delta\phi)$ and is measured by two photodiodes as function of the time delay between the THz and the optical pulses [23,31]. The utilized liquids, are all spectroscopic grade with purity > 99%. The absorption coefficient and refractive index of some of the polar and nonpolar liquids and also the TGG crystal (0.5 mm thick) are given in Fig. 1b.

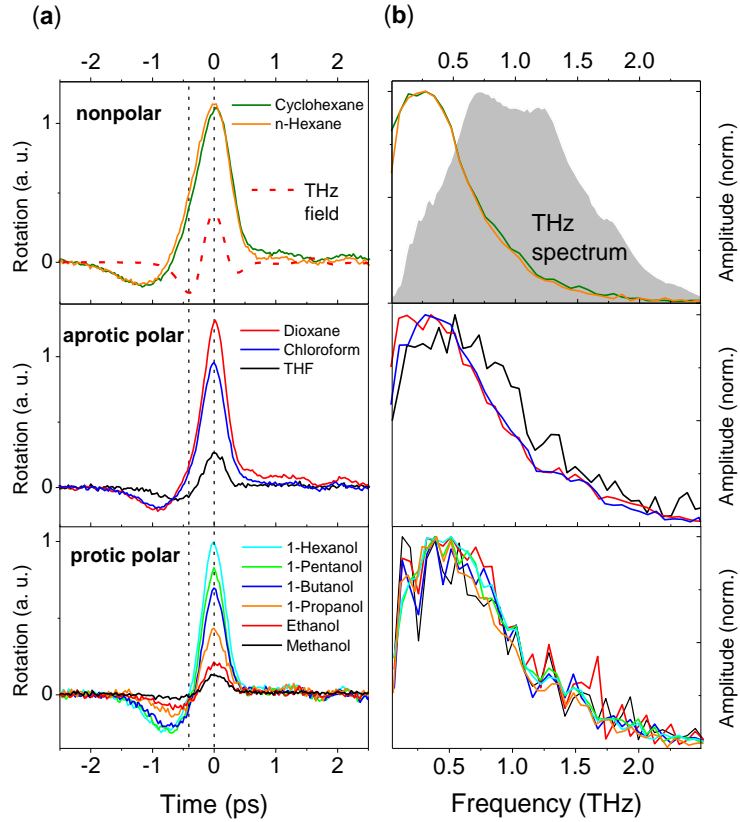

**Figure 2**. a, Faraday rotation of a series of molecular liquids at room temperature: non-polar: cyclohexane (dark green) and n-hexane (red), aprotic polar: dioxane (pink), chloroform (blue), tetrahydrofuran (THF, black) and protic polar: 1-hexanol (cyan), 1-pentanol (green), 1- butanol (blue), 1-propanol (pink), ethanol (red) and methanol (black). The measured signals resemble the THz waveform (dashed line). b, The Fourier spectra of the signals in panel a. The gray area shows the amplitude spectrum of the THz pulse.

Fig. 2a shows the measured Faraday rotation $\Delta\theta$ of the optical probe pulse for all liquids. For comparison, the THz pulse is given by a dashed red line. All signals are bipolar and to a large extent resemble the THz waveform. The Fourier spectrum of the signals is given in Fig. 2b. The gray area shows the spectrum of the THz pulse. The results clearly show, the THz induced rotation of the probe polarization in the sub-picosecond time scale. Note also that the observed rotation signal is independent of the probe polarization. As shown in Fig. S2, identical signals are obtained for $P$ and $S$ polarized optical probe pulses. We further determine the dependence of $\Delta\theta$ on the THz field, by driving the liquids with 180° phase-shifted THz pulses. As shown in Fig. 3, $\Delta\theta$ signals of n-hexane and propanol flip sign when flipping the THz field polarity, indicating that $\Delta\theta$ scales linearly with the THz field. The fluence dependence result, shown in the inset of Fig. 3b, corroborates the latter finding, too.

Interestingly, for water both the ellipticity [32] and the rotation signals are bipolar. However, as shown in Fig. S1, while the ellipticity signal remains intact, the rotation signal flips sign by shifting the phase of the THz pulse by 180°, declaring the different nature of the two effects. The unipolar THz electric-field-induced ellipticity of the rest of the liquids have been published elsewhere [18,33].

Additionally, we compared the polarization rotation of liquids with that of a standard Faraday medium with very high magneto-optical Verdet constant [25]. As shown in Fig. 4, the TGG Faraday rotation has a bipolar shape [34] but, because of a large velocity mismatch between THz and optical pulses, the measured response is broadened. This broadening can be accurately simulated by $\Delta\theta = \frac{2\pi}{\lambda} \int_0^L dx\, \Delta n(0, t + \beta x)$, where the inverse-velocity mismatch $\beta = v_{opt}^{-1} - v_{THz}^{-1}$ quantifies the temporal walk-off of the pump and probe pulses per propagation length [35].

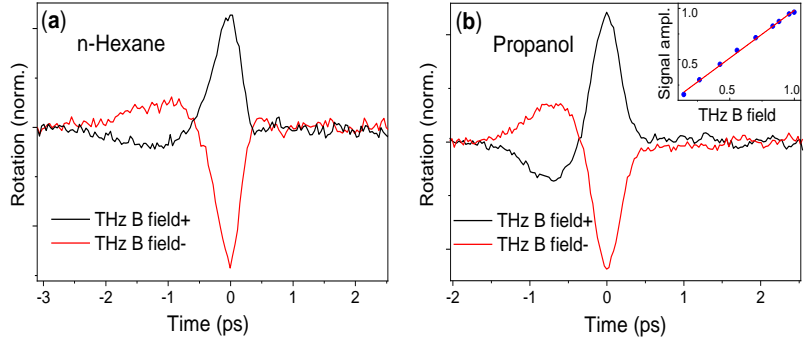

**Figure 3.** Faraday rotation of n-Hexane (a) and propanol (b) scales linearly with the applied THz magnetic-field. In both samples the measured signal flips sign, by 180° phase shift of the THz magnetic-field. Inset in panel b shows the fluence dependence of the peak-to-peak amplitude of the rotation signal of propanol to the THz magnetic-field.

To elucidate the nature of the light-matter interaction causing the rotation signal of liquids, we refer to the electric Hall effect by which the deflection of an electric current in the presence of an external magnetic-field is explained. In a simple classical picture, the Lorentz force $e\mathbf{v} \times \mathbf{B_0}$ bends the electric current out of the plane of a static magnetic-field $\mathbf{B_0}$ and the velocity vector $\mathbf{v}$ of the electric charge $\mathbf{e}$. Interestingly, in its optical analog i.e. the optical Hall effect [36], the bending of the dielectric displacement -due to the Lorentz force- is monitored by the rotation of the light polarization passing through or reflected from the dielectric medium.

In our molecular counterpart of the electric and the optical Hall effects, the THz magnetic-field deflects an optically induced instantaneous electric current in molecules, resulting in an optical polarization rotation. In the optical Hall configuration, the directions of the magnetic-field and the optical wave-vector are parallel, whereas in our experiment we deviate from this scheme by an angle of ~40° between the two EM pulses to attain high time resolution. As a result, the dynamics up to ~2 THz can be resolved, see Fig. 2b. Accordingly, the Lorentz force is proportional to the projected component of the THz magnetic-field into the direction of the optical pulse propagation, see Fig. 1a.

Upon propagation of an optical pulse in a dielectric medium, an instantaneous polarization $\mathbf{P_e}(t) = \alpha \mathbf{E_{opt}}(t)$ is induced which scales with the polarizability $\alpha$ and electric-field $\mathbf{E_{opt}}(t)$ of the optical pulse. For simplicity, we assume that the molecules are linear rotors with their largest polarizability tensor element along the symmetry axis of molecules, which is fairly accurate for the long chain alcohols and n-alkanes [43]. The optical-field-induced polarization can be seen as a molecular current [37], given by the time derivative of $\mathbf{P_e}(t)$

$$\mathbf{J_e}(t) = \partial_t \mathbf{P_e}(t). \tag{2}$$

A component of the THz magnetic-field $\mathbf{B_{THz}}(t)$, parallel to the propagation direction of the optical pulse, exerts a Lorentz force to the electric current and bends the current $\mathbf{J_e}(t)$ in the plane perpendicular to the optical pulse wave vector. The deflected component of the electric current, namely the Hall current reads

$$\mathbf{J_H}(t) = \mathbf{J_e}(t) \times \mathbf{B_{THz}}(t) \sin\gamma = \partial_t \mathbf{P_e}(t) \times \mathbf{B_{THz}}(t) \sin\gamma. \tag{3}$$

Now the bended current, $\mathbf{J_e}(t) + \mathbf{J_H}(t)$ is the source of the electromagnetic radiation whose polarization is rotated by angle $\theta$ relative to the incoming optical pulse. For the opted configuration, the optical pulse propagates with an angle $\gamma$ relative to the THz propagation path, hence $\sin\gamma$ in Eq. 3 accounts for the projection of the THz magnetic-field into the propagation direction of the optical pulse. As detailed in the Appendix A, under the influence of the THz magnetic-field and after propagation of an effective distance $\mathbf{L_{eff}}$ [38], the left and right circularly polarized components of the optical pulse undergo

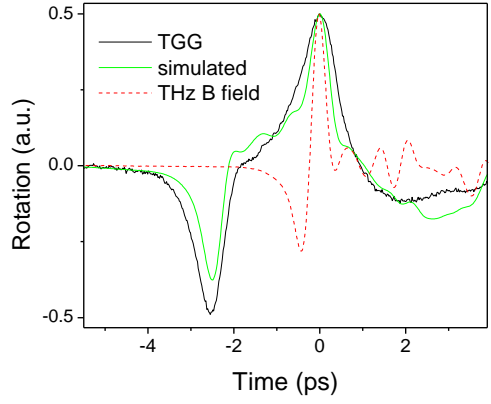

**Figure 4.** The magneto-optic Faraday rotation of a terbium gallium garnet crystal and the simulated signal in which the velocity mismatch between the THz and optical pulses is taken into account (see text). The dashed line shows the THz magnetic-field.

different phase shifts, thus the optical pulse acquires a polarization rotation

$$\Delta\theta(t) = \frac{\pi L}{\lambda}(n_- - n_+) = R_0 L_{eff}\, \alpha^2 B_{THz}(t) \sin\gamma \tag{4}$$

where $R_0 = \omega^2/2n_0 ec$. It is interesting to note that, the rotation signal is related to the molecular property via its electric polarizability and to the optical pulse via its frequency, with no contribution of the optical electric-field amplitude $\mathbf{E_{opt}}$. As detailed in the Appendix A, in the weak interaction regime, $n_L$ and $n_R$ are obtained independent of $\mathbf{E_{opt}}$, resembling the Hall angle in the classical Hall effect [45] in which it scales with the product of the magnetic-field and the mobility of charge carriers.

Eq. 4 indeed explains our main result, a response that scales linearly with the THz magnetic-field $\mathbf{B_{THz}}(t)$ and quadratically with the molecular polarizability. We show the soundness of the latter proportionality by plotting the $\Delta\theta$ signal amplitude versus the square of the average molecular polarizability of liquids. As shown in Fig. 5, there is a linear dependence of the square of the molecular polarizability and the measured rotation signals.

Finally, the Faraday rotation of highly polarizable molecular liquids such as n-hexane and 1-hexanol exceeds that of TGG, more than two folds. Furthermore, TGG has a strong THz absorption and large refractive index, see Fig. 1b, the latter of which causes the large velocity mismatch between THz and optical pulses, hence lowers the effective temporal resolution of the effect. In contrast, in nonpolar liquids, the THz absorption and the velocity mismatch between the two pulses are negligible. Likewise, in long chain alcohols the THz absorption and the velocity mismatch decrease with increasing the alcohol chain length [39].

*Competing nonlinear processes.* In the above description of the THz magnetic-field-induced Faraday rotation in liquids, we considered the optically induced polarization as the dominant contribution to the Hall current (see Eq. 2 and Eq. 3). However, the THz electric-field also contributes to build polarization in liquids, thereby its impact appeals for a careful consideration. While, the first order polarization $\mathbf{P}^{(1)} = \chi^{(1)}\mathbf{E_{THz}}$ has no impact on the optical probe polarization; higher order polarizations have [40]. In the employed THz pump-optical probe experiment, the dynamic THz Kerr effect (TKE) is the expected dominant third-order nonlinear effect from the bulk of liquids [17,18] however, for the following reasons this effect is drastically reduced in our experiment. (i) The selected set of organic liquids possesses relatively small polarizability anisotropy $\Delta\alpha$. The TKE response scales with $\Delta\alpha^2$ in nonpolar liquids and with $\Delta\alpha$ in strongly polar systems, the small $\Delta\alpha$ of the selected liquids directly influences the amplitude of the TKE signal. (ii) In the optimum TKE configurations, the THz pump and the optical probe pulses propagate collinearly through the sample. The angle $\gamma$ in Fig. 1a, reduces the amplitude of the TKE signal. (iii) The dynamic Kerr effect in liquids, induces ellipticity in the optical probe pulse [18] thereby, measuring the rotation instead of ellipticity reduces the Kerr effect contribution on the $\Delta\theta(t)$ signal.

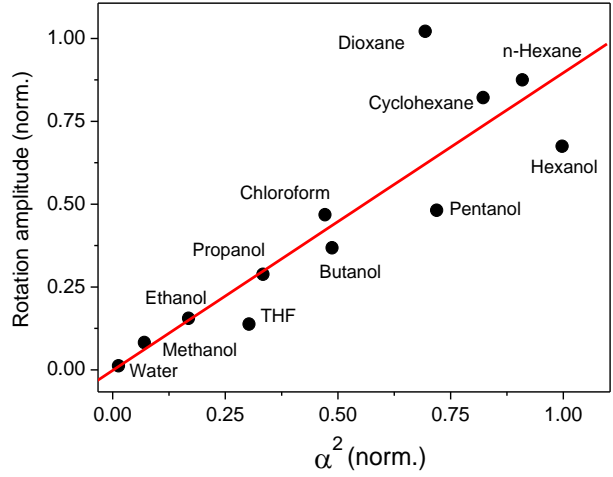

**Figure 5.** Peak-to-peak amplitude of the Faraday rotation signal of liquids is plotted versus the square of the average molecular polarizability of liquids, taken from Refs: water [41], cyclohexane [42], chloroform, methanol, ethanol, propanol, butanol, pentanol, hexanol, *n*-hexane, dioxane [43], tetrahydrofuran (THF) [44].

Although the second order nonlinear processes are forbidden in the bulk of materials with inversion symmetry, they are allowed on their interfaces, where the inversion symmetry is broken [46]. To examine the contribution of the surface nonlinear response $\mathbf{P}^{(2)} = \chi^{(2)} \mathbf{E}_{\mathbf{THz}} \mathbf{E}_{\mathbf{optical}}$ on the rotation signals, which would also scale linearly with the $\mathbf{E}_{\mathbf{THz}}$, we intersect the THz pump and the optical probe pulses in the depth of a liquid. As shown in Fig. S3, the mixing position of the two pulses has no impact on the observed rotation signal, dismissing the $\mathbf{P}^{(2)}$ contribution on the rotation signals.

Does the THz-magnetic-field act on the electric-field-induced instantaneous polarization $\mathbf{P}_e(t) = \alpha\, \mathbf{E}_{\mathbf{THz}}(t)$ of the same pulse? If this effect occurs, its Hall current and accordingly its $\Delta\theta(t)$ signal would scale quadratically with the THz electric-field, as $\mathbf{J}_{\mathbf{H}} = \alpha\, \partial_t \mathbf{E}_{\mathbf{THz}}(t) \times \mathbf{B}_{\mathbf{THz}}(t)$ and $\mathbf{B}_{\mathbf{THz}}(t) = \mathbf{E}_{\mathbf{THz}}(t)/c$. However, the results in Fig. 3 show a linear relation of $\Delta\theta(t)$ and the THz field. Although, the measured $\Delta\theta(t)$ signals in Fig. 2 do not support observation of this effect, it might still exist but overlaid under the stronger Faraday rotation signal from $\mathbf{P}_e(t) = \alpha\, \mathbf{E}_{\mathbf{opt}}(t)$.

In conclusion, THz magnetic-field-induced Faraday rotation in molecular liquids has been resolved on ultrafast timescales. The mechanism of the optical pulse polarization rotation is explained by the deflection of an optically induced instantaneous polarization in the presence of the THz magnetic-field, in a manner analogous to the electric Hall effect in conducting materials. The observed effect scales linearly with the THz magnetic-field and quadratically with the molecular polarizability. Highly polarizable, but lowly THz absorbing liquids such as long chain alcohols and alkanes show the largest effect. The ability of the transient molecular Hall effect to selectively resolve the electron transfer in solutions, paves the way for the disentanglement of electronic versus nuclear dynamics in complex molecular systems. Moreover, the diverse physical and chemical properties of liquids may offer alternative solutions to the material incompatibility in traditional application of the Faraday effect in solid-state photonic devices [47].

APPENDIX A

*Modeling the dynamic molecular Hall effect.* To model the Faraday rotation in molecular liquids, we consider the propagation of a linearly polarized optical pulse through a dielectric medium in the presence of a THz magnetic field. We assume that the optical pulse is linearly polarized in the $x - y$ plane and propagates orthogonal to the propagation path of the THz pulse, while the THz magnetic field points into the $z$ direction (see Fig. 1). As discussed in the main text, see also Eq. 2 and Eq. 3, the THz magnetic field bends the optically induced instantaneous polarization in the medium. The latter bended component of the polarization can be seen as a radiation source in the wave equation to explain the rotation of the optical pulse polarization rotation:

$$\nabla^2 E_\pm - \frac{1}{c^2}\frac{\partial^2 E_\pm}{\partial t^2} = -\frac{1}{c^2\varepsilon_0}\frac{\partial^2 P_\pm}{\partial t^2}. \tag{A1}$$

Note that, as the polarization rotation of the optical pulse scales with the phase difference of its left and right handed circularly polarized components, we decompose the electric field and the induced polarization of the liquid into their circular components $E_\pm = E_x \pm iE_y$ and $P_\pm = P_x \pm iP_y$ where, $P_\pm$ is the deflected component of an optically induced instantaneous electric dipole moment.

To obtain $P_\pm$, we solve the equation of motion of the bound electrons under the influence a Lorentz force. For simplicity, we ignore the local field effects and assume no absorption of the optical pulse in the medium. Moreover, as the temporal duration of the THz pulse is about 50 times longer than that of the optical pulse, we assume that the THz magnetic field has a constant amplitude in its propagation path in the medium. The equation of motion for electrons under the influence of the Lorentz force reads as

$$\ddot{r} + \omega_0^2 r = -\frac{e}{m}(E + \dot{r} \times B) \tag{A2}$$

where B is the magnetic field of the driving THz EM wave, E is the electric field of the optical probe pulse and $e$ ($m$) is the charge (mass) of an electron. For the selected propagation coordinate, Eq. A2 can be decomposed into its components $\ddot{x} + \frac{e}{m}B_z\dot{y} + \omega_0^2 x = -\frac{e}{m}E_x$ and $\ddot{y} - \frac{e}{m}B_z\dot{x} + \omega_0^2 y = -\frac{e}{m}E_y$ and by defining $R_\pm = x \pm iy$, it reads [48,49]

$$\ddot{R}_\pm + \frac{e}{m}B_z\dot{R}_\pm + \omega_0^2 R_\pm = -\frac{e}{m}E_\pm \tag{A3}$$

The steady-state solution under a harmonic electric field $E_\pm = E_0 \exp[\pm i(\omega t - k_\pm z)]$ yields

$$R_\pm = \frac{-\frac{e}{m}E_\pm}{(\omega_0^2 - \omega^2) \pm \Omega\omega} \tag{A4}$$

where $\Omega = \frac{e}{m}B_z$. As a result, the induced electric polarization $P_\pm = -NeR_\pm$ for a medium with electron number density $N$, is given by

$$P_\pm = \frac{N\frac{e^2}{m}E_\pm}{(\omega_0^2 - \omega^2) \pm \Omega\omega} \tag{A5}$$

Eq. A5 in connection with Eq. A1 already reveals the dielectric displacement of the medium in the presence of the THz magnetic field, as $D_\pm = \varepsilon_0 E_\pm + P_\pm = \varepsilon_\pm E_\pm$. Accordingly, the corresponding refractive indices for the left and right handed polarization of the optical pulse can be determined

$$n_\pm^2 = \frac{\varepsilon_\pm}{\varepsilon_0} = 1 + \frac{\frac{e^2}{m}}{(\omega_0^2 - \omega^2) \pm \Omega\omega} \tag{A6}$$

The birefringence given in Eq. A7 leads to the rotation of the plane of the probe polarization by an angle $\theta$ after traversing the sample with thickness L

$$\theta = \frac{\pi L}{\lambda}(n_- - n_+) \tag{A7}$$

Using Eq. A7 and the fact that the expected Faraday rotation is small, such that $n_-^2 - n_+^2 \approx 2n_0(n_- - n_+)$ we obtain

$$\theta = \frac{\pi L}{n_0 \lambda} \frac{\frac{e^2}{m}\Omega\omega}{(\omega_0^2 - \omega^2)^2} \tag{A8}$$

Here, we have assumed that optical pulse frequency $\omega$ is far off the resonant frequency $\omega_0$ of the dielectric medium and for ~1T THz magnetic field $\Omega = \frac{e}{m}B_z \approx 2\pi \times 10$ GHz $\ll \omega$. Finally, using the electric polarizability definition $\alpha = \frac{e^2}{m}(\omega_0^2 - \omega^2)^{-1}$ [ **Error! Bookmark not defined.**], Eq. A8 take the form

$$\theta = \frac{\omega^2}{2n_0 c e}\alpha^2 L_{\text{eff}} B. \tag{A9}$$

Note that $L_{\text{eff}}$ is the effective length of the sample, after considering the geometrical constrains and the THz absorption by the liquids. Moreover, so far we have assumed that THz and optical pulses propagate in a crossing angle of $\gamma = 90°$; however, in the experiment it is set to $\gamma \approx 40°$. The latter difference brings a $\sin\gamma$ factor in to the equation. Accordingly, the Faraday rotation in a dielectric medium with polarizability $\alpha$, refractive index $n_0$ reads

$$\theta = \frac{\omega^2}{2n_0 ec}\alpha^2 L_{\text{eff}} B_{\text{THz}}(t) \sin\gamma. \tag{A10}$$

ACKNOWLEDGMENT

We wish to thank Dr. Lukáš Nádvorník for fruitful discussions. VB acknowledges funding by the H2020 FET-OPEN project ASPIN (grant no. 766566).